\documentclass{article}
\begin{document}

\begin{center}
\centerline{\large \bf On the physical nature of the electromagnetic 
induced}
\centerline{\large\bf transparency effect.}
\end{center}

\vspace{3 pt}
\centerline{\sl V.A.Kuz'menko}
\vspace{5 pt}
\centerline{\small \it Troitsk Institute for Innovation and Fusion 
Research,}
\centerline{\small \it Troitsk, Moscow region, 142190, Russian 
Federation.}
\vspace{5 pt}
\begin{abstract}

	The origin of the electromagnetic induced transparency (EIT) 
effect is explained not as the vanish of atom-field interaction, but 
as the growing of stimulated emission process due to the efficient four-
photon mixing, which allows the atom to return in the initial state. 
We point out the importance of creation the new mathematical model 
for description the dynamics of optical transitions, which should be 
based on the concept of the time invariance violation in electromagnetic 
interactions.

\vspace{5 pt}
\end{abstract}

\vspace{12 pt}

The Bloch equations had been proposed in 1946 for description the nuclear 
magnetic resonance effect [1]. Now there are the basis for description the 
dynamics of optical transitions [2]. The mathematical model well describes the 
physical phenomena, but a great difficulties appear, when anyone wants to 
understand the physical sense of this descriptions. The rapid adiabatic 
passage (RAP) effect is the most simple and fundamental example of such 
phenomena. When the resonance radiation interacts with the two-level system, 
the so-called periodical Rabi oscillations of the level population take 
place. In this case the situation is quite symmetrical. But if the sweeping 
of resonance conditions appears (for example, the frequency of radiation is 
changed), the full population transfer from the initial level to the opposite 
one takes place. The theory well describes this effect, but it does not give 
physical explanation [3]. This is the surprising situation: the mathematical 
description exists more than fifty years, but any physical explanation of 
effect is absent till now. 

This situation is not an accidental case. It is impossible to explain the 
origin of RAP effect, if we suppose the equality of forward and reversed 
transitions. The restriction to equality of integral cross-section of the 
transitions (equality of the Einstein coefficients) is sufficient for 
explanation the Rabi oscillations. The physical explanation of RAP effect is 
possible only if we suppose some inequality between the forward and reversed 
transitions (in the spectral width, or in the differential cross-section). 
So, the existence of RAP effect is the indirect proof of such inequality.

Moreover, the number years we have quite direct and complete experimental 
proof of inequality of forward and reversed transitions. This proof is 
connected with the existence of the so-called wide component of line in 
absorption spectrum of polyatomic molecules [4]. This physical object has 
unusual properties: the large spectral width of optical transition 
($\sim 150 GHz $) is combined with the long lifetime of the molecule excited 
states. Its allow to measure separately the parameters of forward and reversed 
transitions. The measured spectral width of the reversed transition was less 
than 1 MHz ([5] Fig.5). So, the ratio of spectral width of forward and 
reversed transitions in this case exceeds five orders of magnitude. 
Accordingly, the cross-section of reversed transition was found to be orders 
of magnitude greater, than for the forward one [6]. Such result is a good 
base for explanation the origin of the effects in nonlinear optics [7].

	The goal of this note is to propose the physical explanation of EIT 
and related effects (self-induced transparency [8], optical pulse compression 
in a resonant absorber [9], etc.). The usual explanation of EIT effect is 
based on the conceptions of coherency and interference. There are rather 
indistinct and vague concepts. It is supposed, that "the effect of EIT is 
due to the existence of a coherent superposition of quantum states which 
does not participate in the atom-field interaction ("dark" state) because of 
a quantum interference of the excitation paths" [10]. 

In contrast, the present explanation turns our attention on a stimulated 
emission process. Absorption of energy is a difference between processes of 
excitation and deexcitation of atoms due to stimulated emission. In terms of 
ultranarrow width of the reversed optical transition the stimulated emission 
process may have the difficulty due to the uncontrolled sweeping of resonance 
conditions (for example, as the result of the ac Stark shift of atom levels in 
the laser field [11]). Such difficulty can be overcome through the four-photon 
mixing process, which is strongly facilitated by the presence of powerful 
resonance radiation on the coupling transition of the lambda-coupling scheme. 
As the result, the regeneration of the probe laser radiation takes place and 
the atom returns in the initial state. 

	The published experimental results sufficiently clearly show both the 
dynamics of the probe laser field regeneration [12-14] and the existence of 
the extremely efficient four-photon mixing process in this conditions 
[10,15-17]. The light "stopping" effect in such experiments does not have, 
of course, any connection with the speed of photons in the low presure gas 
mixture. The delay time of the probe laser pulse is the result of the 
multiple stage of four-photon mixing process and characterizes the build up 
time of the superfluorescence emission [12]. 

All processes, which can destroy the four-photon mixing, should lead to 
growing the absorption of the probe beam. For example, the additional driving 
field gives the so-called double dark resonance [18]. In this case only six-
photon mixing can return atom in the initial state. So, this process is less 
efficient, than a four-photon mixing. When the probe laser deturn from the 
resonance in lambda-type coupling scheme, the growing of absorption is 
observed ([12] Fig.1). In this case we obviously have the variant of the 
stimulated Raman population transfer [19]. 

Extremely high efficiency of a four-photon mixing processes in nonlinear 
optics can be explained as the result of some "memory" of atoms and molecules 
about the initial state and their aspiration to return back. This "memory" is 
connected with the inequality of the forward and reversed optical transitions. 
In other words, this is the result of the time invariance violation in 
electromagnetic interactions [20]. In spite of common opinion the inequality 
of forward and reversed transitions is widespread in the optics. There is the 
base of most phenomena in the nonlinear optics. But it usually manifests 
itself only in an indirect way, because of the direct and independent 
measurements of the forward and reversed transitions have some difficulties.

It is important to create the new mathematical model (alternative to the
Bloch equations) for adequate description the dynamics of optical transitions. 
This model should be based on the concept of time invariance violation in 
electromagnetic interactions [21]. The descriptions of physical phenomena with 
such mathematical model will have much more straightforward and clear physical 
sense, than the present day theory descriptions.

\end{document}